\documentclass[twocolumn,article,notitlepage,floatfix,superscriptaddress,nofootinbib]{revtex4-1}

\usepackage{amsmath}
\usepackage{amssymb}
\usepackage{braket}
\usepackage{titlesec}
\usepackage{graphicx}
\usepackage{array}
\usepackage{hyperref}
\usepackage{color}                     
\usepackage{xcolor}
\usepackage[caption=false]{subfig}
\graphicspath{ {./images/} }

\newcommand{\figref}[1]{Figure {\ref{#1}}}
\newcommand{\secref}[1]{Section {\ref{#1}}}

\newcommand{\tableref}[1]{Table {\ref{#1}}}

\renewcommand{\eqref}[1]{Eq. ({\ref{#1}})}
\usepackage{ulem}

\definecolor{applegreen}{rgb}{0.55, 0.71, 0.0}

\begin{document}

\title{Do Quantum Circuit Born Machines Generalize?}

\author{Kaitlin Gili}
\affiliation{University of Oxford, Oxford, United Kingdom OX1 2JD }
\affiliation{Zapata Computing Canada Inc., 325 Front St W, Toronto, ON, Canada M5V 2Y1}
\author{Mohamed Hibat-Allah}
\affiliation{Zapata Computing Canada Inc., 325 Front St W, Toronto, ON, Canada M5V 2Y1}
\affiliation{Vector Institute, MaRS Centre, Toronto, ON, Canada M5G 1M1}
\affiliation{Department of Physics and Astronomy, University of Waterloo, Waterloo, ON, Canada N2L 3G1}
\author{Marta Mauri}
\affiliation{Zapata Computing Canada Inc., 325 Front St W, Toronto, ON, Canada M5V 2Y1}
\author{Chris Ballance}
\affiliation{University of Oxford, Oxford, United Kingdom OX1 2JD }
\author{Alejandro Perdomo-Ortiz}
\email{alejandro@zapatacomputing.com}
\affiliation{Zapata Computing Canada Inc., 325 Front St W, Toronto, ON, Canada M5V 2Y1}

\date{\today} 
\normalem
\begin{abstract}

In recent proposals of quantum circuit models for generative tasks, the discussion about their performance has been limited to their ability to reproduce a known target distribution. For example, expressive model families such as Quantum Circuit Born Machines (QCBMs) have been almost entirely evaluated on their capability to learn a given target distribution with high accuracy. While this aspect may be ideal for some tasks, it limits the scope of a generative model’s assessment to its ability to \emph{memorize} data rather than \emph{generalize}. As a result, there has been little understanding of a model's generalization performance and the relation between such capability and the resource requirements, e.g., the circuit depth and the amount of training data. In this work, we leverage upon a recently proposed generalization evaluation framework to begin addressing this knowledge gap. We first investigate the QCBM's learning process of a cardinality-constrained distribution and see an increase in generalization performance while increasing the circuit depth. In the 12-qubit example presented here, we observe that with as few as $30\%$ of the valid data in the training set, the QCBM exhibits the best generalization performance toward generating unseen and valid data. Lastly, we assess the QCBM's ability to generalize not only to valid samples, but to high-quality bitstrings distributed according to an adequately re-weighted distribution. We see that the QCBM is able to effectively learn the reweighted dataset and generate unseen samples with higher quality than those in the training set. To the best of our knowledge, this is the first work in the literature that presents the QCBM's generalization performance as an integral evaluation metric for quantum generative models, and demonstrates the QCBM's ability to generalize to high-quality, desired novel samples.
\end{abstract}

\maketitle

\section{Introduction}\label{s:intro}

From classical machine learning (ML), we have seen remarkable applications across a wide variety of industries including image classification and generation \cite{huang2018introduction, karras2020analyzing}, language processing \cite{2011Nadkarni}, and constructing complex recommendation systems \cite{Recommender2016}. As such, it has become an active area of research to understand how quantum computers may enhance, or outperform, these classical algorithms. 

In the pursuit of practical quantum advantage on classical data, unsupervised generative modeling tasks stand out as one of the most promising application candidates given their increased complexity compared to supervised ML tasks, and therefore a better target for seeking advantage with near-term quantum computers~\cite{PerdomoOrtiz2017}. Many quantum generative models have been proposed with very little discussion around their learning potential in the context of generalization~\cite{qgen_review}, despite its importance. One of the most popular quantum circuit families for generative tasks, known as Quantum Circuit Born Machines (QCBMs)~\cite{Benedetti2019}, have demonstrated remarkable capabilities in modeling target distributions for both toy and real-world datasets~\cite{Benedetti2019, liu2018differentiable, Zhu2018, Hamilton2018, Zoufal2019, Alcazar2020ClvsQuant, coyle2020generativeFinance, Benedetti2021, rudolph2020generation,Zhu2022}. It has been shown that these models have the ability to express distributions that are difficult for classical probabilistic models~\cite{Coyle2019,du2018expressive, glasser2019, sweke2020learnability, gao2021enhancing}, further motivating an investigation into these models for quantum advantage applications. 

When it comes to generalization, discriminative models have been the primary focus of research - both in the classical and quantum domain \cite{abbas2021effective, Caro_2022, zhang2017understanding}. These type of generalization studies are based on the so-called generalization error, which describes how well the model is able to classify unseen data after learning with a labelled training set. While the metrics for measuring generalization within classical and quantum discriminative tasks are reasonably intuitive and well-established in the field of ML, this is far from being true in the context of unsupervised generative tasks. The generalization behavior that matters in unsupervised generative modeling is defined as the model’s ability to generate new samples from an underlying, unknown probability distribution after training on a finite set of samples. In fact, developing new evaluation metrics is still an active area of research~\cite{gili2022evaluating, alaa2021faithful, borji2021pros, meehan2020non, sajjadi2018assessing, kynkaanniemi2019improved, naeem2020reliable, zhao2018bias}. 

So far in the literature, assessing the quality of these quantum-circuit-based generative models has thus been almost entirely limited to how well they can memorize or reproduce a known target distribution. Despite its intrinsic value for benchmarking purposes, by sticking to a reproducibility/data-copying metric such as minimizing the empirical Kullback-Leibler (KL) divergence or negative log-likelihood (NLL) as the primary method of evaluating a generative model, we are not properly assessing the model’s true ability to generalize, since the optimal solution would correspond to memorizing the training data. A recent work \cite{mmd} specifically claims to focus on a learning theory for quantum generative models from the perspective of generalization. Despite the authors' proof of theoretical generalization bounds for quantum generative models trained via Maximum Mean Discrepancy, their approach (based on Ref.~\cite{dziugaite2015training}) does not explicitly take into account the crucial feature of novelty, which constitutes an essential ingredient to define generalization. The generalization error proposed in \cite{mmd} quantifies the deviation of the empirically optimized probability distribution encoded by a model after training from the best probability distribution the model can represent, given its expressivity and a desired target distribution~\footnote{In practice, the target distribution is unknown, so including it in the model evaluation process is only feasible from a theoretical perspective.}. When this deviation is minimized, the learning performance is maximized. However, if a generative model is simply memorizing data, such deviation will be exactly zero, hence implying that this metric is not sensitive to the novelty of the generated samples. Since the comparison between training set and generated queries is out of the scope of Ref.~\cite{mmd}, the model's true ability to generalize is not fully assessed.

In other words, most proposals fail to conduct a complete investigation as to how the model learns features of complex target distributions from a finite set of training data, and as a result ignore the role that important resource requirements (e.g., circuit depth and amount of training data) play in understanding the model's generalization capabilities in tasks of practical relevance. Recently, theoretical rigorous results were obtained for certain output distributions from local quantum circuits, proving some challenges in achieving a separation advantage with respect to classical models within this family~\cite{Hinsche2022}. While those results focus on worst-case bounds for the entire family of distributions, here we focus on the performance of QCBMs on specific realizations of application-relevant classical distributions.

Recently, a novel evaluation framework has opened the door for investigating generalization in classical and quantum generative models in greater depth \cite{gili2022evaluating}. In this work, we leverage such framework to evaluate, for the first time, the generalization capabilities of quantum-circuit-based models. We present the QCBM's generalization performance as an integral component of its evaluation as a generative model, and its ability to perform well with limited training data.
In Sec.~\ref{s:concept_overview}, we review the key concepts, models, and metrics used in this work. In Sec.~\ref{s:results}, we present the main results from this study.
First, we investigate the model's validity-based generalization performance, i.e., its ability to learn the valid features of non-exhaustive training datasets and generate novel valid samples post training. We evaluate the generalization throughout training, and observe improved performance at each iteration and overall when increasing the number of circuit layers. We then conduct an initial assessment of the QCBM's scarce-data regime by reducing the number of valid data shown to the model during training, and investigate the minimum data requirement needed to achieve high quality generalization performance. Lastly, we investigate the model's quality-based generalization performance, i.e., its ability to generalize not only to valid data, but also to valid data whose average associated cost is less than the cost of appropriately re-weighted training samples. We see that the QCBM is able to effectively learn the artificially reweighted training set and generate unseen samples with high quality. Finally, in Sec.~\ref{s:outlook}, we conclude with some potential future research directions from this work.

\section{Key Concept Review}\label{s:concept_overview}

Prior to presenting our generalization results for QCBMs, we provide a review of key concepts utilized throughout this work, including a basic introduction to unsupervised generative models, the QCBM model family, and the validity-based and quality-based frameworks required to assess generalization. Using these concepts, we introduce the overall algorithm and evaluation scheme utilized to quantify the generalization capabilities of QCBMs along with their resource requirements. A visualization of this explanation is provided in \figref{fig:explanation}.

\subsection{Unsupervised Generative Models}\label{sec2:unsup_gen}

Different to discriminative models in supervised learning, unsupervised generative models aim to learn an underlying, unknown, probability distribution $P(x)$ from a finite set of unlabelled training samples. These networks capture correlations within high-dimensional target distributions, often times having limited access to information, which makes generative modeling a much more difficult task compared to discriminative modeling \cite{gao2021enhancing, gui2020review, ruthotto2021introduction}. Many kinds of generative models have been proposed in the literature, and often come with different architectures, training strategies, and limitations \cite{qgen_review,Bengio-Book}. A few notable model families include Generative Adversarial Networks (GANs) \cite{goodfellow2016nips}, Restricted Boltzmann Machines (RBMs) \cite{RBM_Hinton},
Tensor Network Born Machines (TNBMs) \cite{han2018unsupervised} and Quantum Circuit Born Machines (QCBMs) \cite{Benedetti2019}. While some of these models are able to perform data-driven tasks that require generative learning with distributions with continuous variables, we ultimately restrict our subsequent model definitions to networks that learn distributions with discrete variables, since it has been shown that discrete problems give rise to an unambiguous framework for assessing generalization \cite{zhao2018bias, gili2022evaluating}, and are more appropriate when working with quantum circuit models. Also, we highlight that the discrete nature of the dataset does not prevent the problem instance from being of practical relevance. For example, these discrete tasks appear naturally in constrained combinatorial optimization problems.

Given that we have a discrete problem, we define the unsupervised generative task as one that attempts to learn an unknown target distribution $P(x)$ given only a set of training samples from such distribution. This set constitutes the training dataset $\mathcal{D}_{\text{Train}} = \{{x_{1}, x_{2},...,x_{D}\}}$, where each sample $x_{t}$ is an $N$-dimensional binary vector such that $x_{t} \in \{0,1\}^N$ with $t=1,2,\dots,D$. We denote the probability distribution defined by the training set as $P_{\text{train}}(x)$. Post training, the model is queried to generate data that composes the set $\mathcal{D}_{\text{Gen}}=\{x_{1}, x_{2},..., x_{Q}\}$, where each $x_{q}$ is again an $N$-dimensional bitstring, with $q=1,2,\dots,Q$. A good generative model should learn optimal parameters that make the model a faithful approximation of the original target distribution $P(x)$, implying that the model has the ability to generate data $x_{q}$ from both inside and outside of the training set that are distributed according to $P(x)$. 

We refer to the model's ability to generate data outside of the training set that are still distributed according to the data distribution $P(x)$ as \emph{generalization}. Additionally, we note that the model may generate samples that are not in the training set, but also are not in support of the target distribution. The latter correspond to invalid samples, and we refer to them as noise. The model's ability to distinguish between noisy and valid samples is an important property that should be included in the performance assessment of any classical and quantum generative model, and it is one of the generalization metrics further discussed in \secref{sec2:eval_gen}. 

\subsection{Quantum Circuit Born Machines (QCBMs)}\label{sec2:qcbms}

\begin{figure*}
\includegraphics[width=\linewidth]{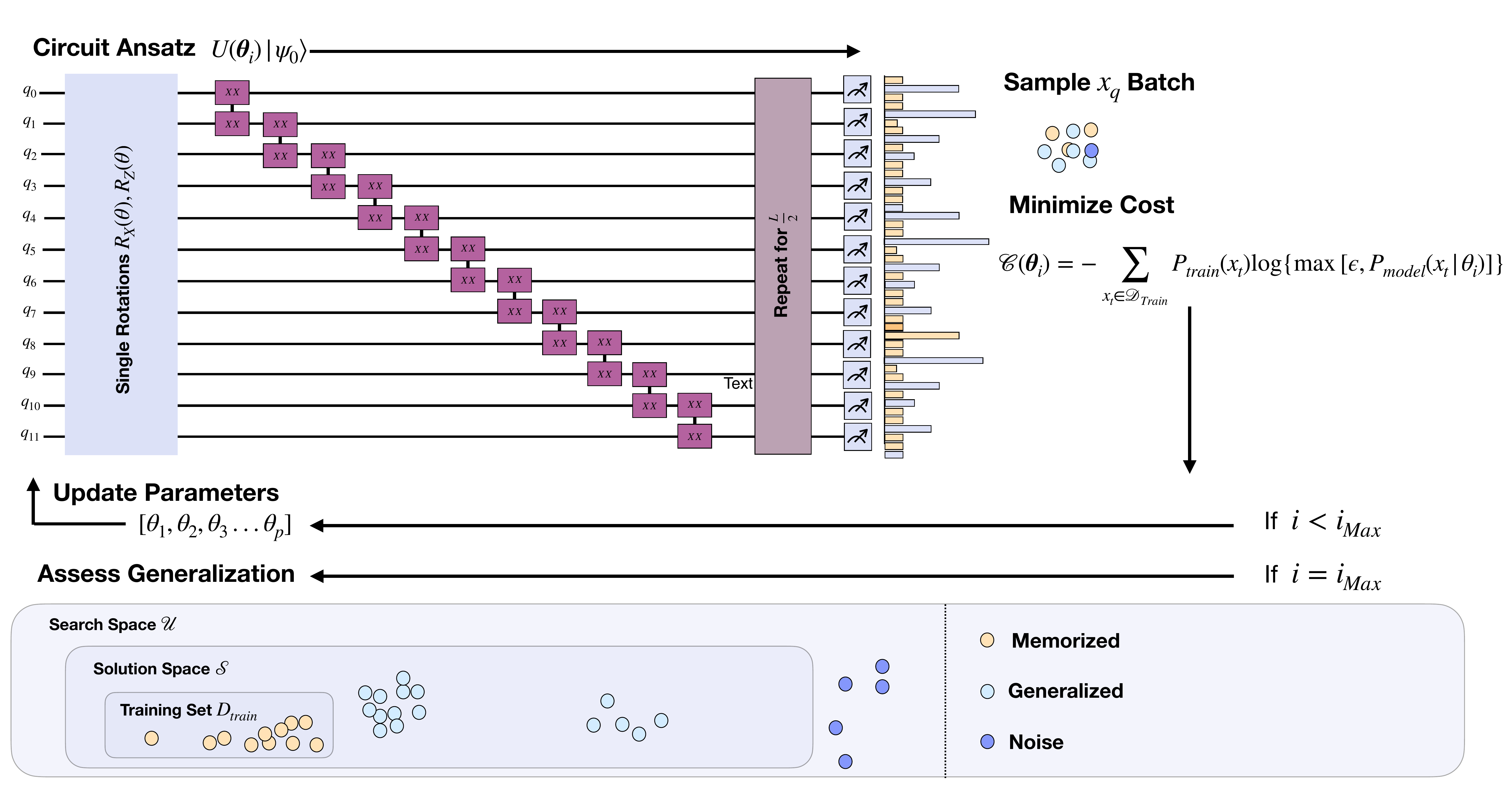}
\caption{\textbf{A visualization of the QCBM algorithm and generalization evaluation scheme.} Starting with randomly initialized parameters, the 12-qubit circuit ansatz with a line topology of parameterized gates for an even number of layers $L$ is executed on a quantum simulator (note that both single-qubit and entangling gates are taken to be parametrized). Measuring in the computational basis provides samples distributed according to the probabilities encoded in the quantum state $|\psi\rangle$ that results from performing the unitary operation $U(\boldsymbol{\theta})$ on an initial quantum state $|\psi_{0}\rangle$. Iterative training is implemented up to a number of $i_{\text{Max}}$ iterations in order to optimize the circuit parameters via minimization of the cost function $\mathcal{C}(\boldsymbol{\theta}_i)$. A post training sample-based evaluation scheme is conducted in order to assess the generalization capabilities of the QCBM. From a batch of sampled queries, each query $x_{q}$ can be categorized as a memorized (yellow), generalized (light blue), or noise (dark blue) count. Given the number of each type of count, the Fidelity, Rate, and Coverage metrics $(F, R,C)$ , or their equivalent normalized version, can be computed for the batch of samples, quantifying the generalization capabilities of the model.
}
\label{fig:explanation}
\end{figure*}

In the literature, QCBMs are one of the most popular quantum generative modeling families due to their highly expressive power~\cite{du2018expressive} and the ability to perform direct sampling from the circuit as opposed to RBMs that require a costly Gibbs sampling. This model family takes advantage of the Born rule of quantum mechanics to sample from a quantum state $|\psi\rangle$ learned via training of a Parameterized Quantum Circuit (PQC) unitary $U(\boldsymbol{\theta})$, where $\boldsymbol{\theta}$ is the vector of parameters for all of the single and entangling gates in the circuit. While alternative ansatz connectivities may be implemented, we showcase a line topology in \figref{fig:explanation} as an appropriate choice for training circuits with very large depths. As the number of entangling gates scales linearly in the number of qubits for each layer, one can increase the number of layers with fewer number of parameters compared to other topologies. Note that alternative topologies, such as the all-to-all entangling connectivity available in ion-trap devices ~\cite{Benedetti2019,Zhu2018,Zhu2022,Rad2022}, may be better for alternative tasks when the circuit depth can remain low. For a total number of layers $L$ in the line circuit ansatz, each layer alternates between parameterized single-qubit rotation gate and multi-qubit entangling gate sequences until the final layer is reached. Each single-qubit gate sequence consists of an appropriate combination of Pauli $X$ and Pauli $Z$ rotations, $R_{X}(\theta)$ and $R_{Z}(\theta)$ respectively on each of the qubits, with $R_{m}(\theta) = \exp{(\frac{-i \theta \sigma_{m}}{2}})$. After each single-qubit gate layer, the entangling layer containing parameterized $XX$ couplers between nearest neighbour qubits is executed, forming a line structure or an all-to-all connectivity as used in some of the cases here. Other topologies can also be explored (see e.g., Ref.~\cite{Zhu2018}).

To minimize the number of variational parameters of the circuit and favor its trainability without sacrificing its expressive power, we can utilize the following strategy to carefully design the single-qubit layers~\cite{Benedetti2019}, for an even $L$. In the first single-qubit gate sequence, we choose to decompose the arbitrary single-qubit transformation as $R_Z(\theta_1)$ $R_X(\theta_2)$ $R_Z(\theta_3)$. Since our initial state is $|00 \cdots 0>$, the first sequence of $R_Z(\theta_1)$ only adds a global phase to the quantum state, which is irrelevant since it will get washed out once we consider the Born's probabilities. Therefore, we can remove this first sequence of $R_Z(\theta_1)$ on each qubit without reducing the circuit's expressive power. For the next single-qubit sequences after the very first one,  we propose a decomposition of the form $R_X(\theta_1)$ $R_Z(\theta_2)$ $R_X(\theta_3)$: it can be seen that the commutation of $R_X$ with $XX$ would lead to ``collapsing" one sequence of $R_X$ from the single-qubit sequence right before the entangling layer into the one right after it. Leveraging this commutation-and-collapse trick, all the single-qubit sequences after the first can also be reduced to $2N$ gates, except for the last one that has $3N$ gates as the final $R_X$ doesn't have any following rotation to collapse into. For $N$ qubits and an even $L$, this ansatz choice gives a total number of parameters $P = (3L / 2 + 1)N - (L / 2)$. When $L = 2$, the parameter count is simply given by $P = 3N - 1$ as there are only $2N$ single qubit gates in the first single-qubit sequence, as previously explained, followed by $N-1$ parametrized XX gates.


During each training iteration $i$, up to $i_{\text{Max}}$, the quantum circuit is simulated or run on quantum hardware with the current iteration parameter values $\boldsymbol{\theta}_{i}$, and is then queried to generate samples $x_{q}$ according to the following model distribution \cite{Benedetti2019}: 
\begin{equation}\label{born_rule}
P_{\text{model}}(x_{q}| \theta_i) = |\langle x_{q} | \psi ( \theta_i)\rangle|^2. 
\end{equation}
The generated samples are input into a classical cost function that measures the distance between the model output samples $x_q$ and the training samples $x_t$ taken from the underlying target distribution. Typical cost functions include the negative log likelihood (NLL) and the Maximum Mean Discrepancy (MMD) loss \cite{liu2018differentiable}. In this work, we utilize the NLL cost function at each iteration defined as: 
\begin{equation}\label{nll}
\mathcal{C}(\boldsymbol{\theta}_{i}) = - \sum_{x_t \in \mathcal{D}_{\text{Train}}} P_{\text{train}}(x_t)\log\{\max{[\epsilon, P_{\text{model}}(x_t | \theta_i )]}\}, 
\end{equation}
where $\epsilon = 10^{-8}$ mitigates the singularity that occurs when $P_{\text{model}}(x_t | \theta_i) = 0$.
A limitation of this cost function is that it requires explicit access to $P_{\text{model}}(x_q)$, where we are only able to approximate this value with a finite number of queries $Q$ taken from the trained model. Thus, this cost function is prone to the curse of dimensionality; it becomes challenging to approximate  when scaling to larger data dimensions.
Once computed, the cost is utilized in a gradient-based or gradient-free optimization scheme that updates the parameter values and feeds them back into the circuit for the next iteration. After a specified number of iterations $i_{\text{Max}}$ (once a cost-threshold $\alpha$ is reached or some other convergence criterion is satisfied), the trained model can be queried and its output is used to evaluate its generalization capabilities.

There have been several proposals to mitigate the singularities and scalability issues arising from the use of the NLL (or, equivalently, the KL Divergence). The first proposal to mitigate this was to use the Maximum Mean Discrepancy (MMD) cost function \cite{liu2018differentiable}. Other papers have proposed other types of divergences, such as the Sinkhorn divergence and the Stein discrepancy \cite{Coyle2019}, and other f-divergences \cite{f_diverge}. Although changing the cost function might mostly help to reduce significantly the resources needed to estimate the cost function itself from the quantum device's samples, this might not suffice to address other trainability issues such as the presence of barren plateaus. Recently, we proposed a synergistic approach which leverages tensor networks' solutions to the problem, and use this approximation to initialize the training of the QCBMs (\cite{rudolph2022synergistic} and \cite{rudolph2022MPSdecomposition}). This strategy was shown not to exhibit such exponential vanishing gradients, and therefore it is expected to successfully scale to much larger problem instances. For the number of qubits studied here, we observed excellent performance via training under the NLL cost function.

\subsection{Generalization Metrics}\label{sec2:eval_gen}

Evaluating unsupervised generative models has remained a challenging task in both the classical and quantum regime. Oftentimes, memorization is the center of this assessment. In the search for evaluating the model's true learning power over its ability to data-copy, one can aim to measure the model's generalization capability. However, this has proven to be a difficult task on its own \cite{qgen_review}. Prominent metrics such as Precision $p$ and Recall $r$ \cite{sajjadi2018assessing, simon2019revisiting, kynkaanniemi2019improved} aim to assess the quality of the generated data. However, these two numbers lack specific information regarding novelty since they include samples that are part of the training set. For example, one might obtain a perfect precision with samples coming exclusively from the training set. As mentioned in \secref{sec2:unsup_gen}, generalization is the phenomenon that occurs when the model is able to produce samples outside of the training set $\mathcal{D}_{\text{train}}$ that are still distributed according to the target distribution $P(x)$. Evaluating generalization thus requires metrics that capture this capability, and the typical methods utilized in literature for measuring the quality of QCBMs, such as the KL Divergence \cite{Benedetti2019,Zhu2018} fail in this regard. This metric will return a perfect score if $P_{\text{model}} = P_{\text{train}}$, and does not provide specific information regarding the model's ability to generalize to $P(x)$ from a finite set of training data. Furthermore, we note that one may split the training set into two, and utilize one as a test set for measuring the KL Divergence with respect to the model output. However, this measure only provides a single perspective as opposed to the detailed picture of the model's generalization capabilities provided by our metrics described next.

In this work, we leverage upon a recently proposed model-agnostic and sample-based framework \cite{gili2022evaluating} that places novelty at the forefront of the evaluation to assess the learning capabilities of QCBMs. 
The first step proposed in the framework evaluates the model's validity-based generalization, and it entails a few requirements. The first is that the target distribution $P(x)$ must describe samples that exist in a \emph{valid} solution space $\mathcal{S}$ that is a subset of a given search space $\mathcal{U}$ of $2^N$ discrete states. The solution space is defined by some desired feature, where bitstrings that have this feature are considered valid, and are considered noise otherwise. However, only valid bitstrings that are unseen (i.e., not in the training set) are considered to be generalized samples. Generated bitstrings that can be found in the training set are considered to be memorized samples (see Figure~.\ref{fig:explanation}). As a consequence, it is important that the training set is non-exhaustive so that there is room for generalization. 

By drawing a specified number of queries $Q$, a model is assessed for its validity-based generalization capabilities by computing the \emph{Fidelity (F)}, \emph{Rate (R)}, and \emph{Coverage (C)} metrics' values, defined as: 

\begin{equation}\label{fidelity}
F = \frac {|\mathcal{G}_{\text{sol}}|}{|\mathcal{G}_{\text{new}}|},
\end{equation}

\begin{equation}\label{rate}
R = \frac {|\mathcal{G}_{\text{sol}}|} {Q},
\end{equation}

\begin{equation}\label{coverage}
C = \frac {|g_{\text{sol}}|} {|\mathcal{S}| - D}.
\end{equation}
In the formulas above, $\mathcal{G}_{\text{new}}$ is the multi-subset of unseen queries (noisy or valid), $\mathcal{G}_{\text{sol}}$ is the multi-subset of unseen and valid queries, and $g_{\text{sol}}$ is the subset of unique unseen and valid queries. We note that $F$ and $R$ do not require a priori knowledge of $P(x)$ in order to be computed, while a limitation of $C$ is that it requires knowledge of the size of the solution space $|\mathcal{S}|$. 

Each of the $F, R, C$ values provides unique information about the model's generalization capabilities, and therefore about its learning capabilities. Conceptually, the fidelity describes how well the model can generalize to valid samples rather than produce noise; the rate describes the frequency at which the model generates unseen valid samples; and the coverage describes the portion of the unseen valid space the model is able to learn. Altogether, these metrics provide a 3D picture of the model's capability to learn valid features in the dataset and produce novel valid samples. 

In addition to the generalization metrics introduced above, we also define the typical, aforementioned, precision metric $p$ and the pre-generalization exploration metric $E$. While these values do not quantify generalization performance directly, they do add information when compared alongside the generalization metrics. $p$ is computed as follows: 
\begin{equation}\label{precision}
p = \frac {|\mathcal{G}_{\text{train}}|+|\mathcal{G}_{\text{sol}}|} {Q},
\end{equation}
where $\mathcal{G}_{\text{train}}$ is the number of queries that were memorized from the training set. $E$ is computed as follows: 
\begin{equation}\label{exploration}
E = \frac {|\mathcal{G}_{\text{new}}|} {Q},
\end{equation}
where the quantity shows the fraction of unseen samples that were queried from the model. 

However, we note that when one is varying the size of the training set $D$, which can be controlled by increasing or decreasing the variable $\epsilon$ such that $D = \epsilon |\mathcal{S}|$, the metric computations may be individually affected as discussed below. 

When computing the coverage metric across $\epsilon$, we propose to use a normalized coverage $\tilde{C} = C / \overline{C}$, where $\overline{C}$ is taken to be the expected value of coverage, computed by: 

\begin{equation}\label{exp_coverage}
\overline{C} = 1 - \left (1 - \frac{1}{|\mathcal{S}| (1 - \epsilon)} \right )^Q.
\end{equation}
This estimator, factoring in $\epsilon$ and the number of queries $Q$, indicates which coverage $C$ one should expect when the generative model has perfectly learned the target distribution and generates samples accordingly \cite{gili2022evaluating}. When the number of queries is much smaller compared to the number of unseen solutions, i.e., $Q \ll |S|(1-\epsilon)$, we can approximate $\overline{C}$ as $\frac{Q}{|S|(1-\epsilon)}$. In this regime, the normalized coverage can be estimated as
\begin{equation}
    \tilde{C} \approx \frac{|g_{\text{sol}}|}{Q}.
    \label{coverage_asymptotic}
\end{equation}
Interestingly, this expression does not need an explicit knowledge of $|S|$.

We additionally propose a normalized rate $\tilde{R} = R / \overline{R}$, where: 
\begin{equation}\label{exp_rate}
\overline{R} = 1 - \epsilon.
\end{equation}
The latter is the rate one should expect if the target distribution is learned perfectly. More specifically, the probability that a generated query is valid and unseen (i.e, rate) corresponds to the portion of the valid space that is unseen for a given $\epsilon$, i.e., $(\mathcal{S}-D)/\mathcal{S}$.

For the fidelity $F$, we note that proposing a normalized $F$ is non-trivial as this metric does not have an ideal value that depends on $\epsilon$. If the target distribution is learned exactly, then we should see $F = 1$ independently of $\epsilon$. But there is a non-trivial dependence on $\epsilon$ for any other model which is not perfect. To illustrate this dependence, imagine one obtains the same model after training under two different values of $\epsilon$. These two models will yield, for example, the same values for the precision, since $p$ does not depend on the size of the training set but only on the probability of a query being in the valid sector. Since a larger $\epsilon$ implies that the portion of seen and valid data (the training set) is larger, this automatically implies that the sector of unseen and valid data is smaller and therefore the probability for a query to land there is smaller. Since the probability of landing in noise (unseen but invalid) remains the same, but the fraction of unseen and valid which appears in both the numerator and denominator of $F$ changes, this yields different values for $F$ for the same model. We leave the development of a normalized fidelity to future work, and highlight that this comment only becomes relevant when comparing across different $\epsilon$ values. When comparing models with the same $\epsilon$, or assessing individual models, the metrics can be still used for accessing generalization.

Altogether, we are able to utilize these metrics in the first part of the framework to obtain a well-rounded picture of the QCBM's validity-based generalization capabilities. 

The second step proposed in the framework, introduced in \cite{gili2022evaluating} as the quality-based generalization evaluation, requires each sample from the target distribution $P(x)$ to have an associated cost, so that all the training samples can be ranked by this value. The discrete training distribution is no longer expected to be uniform probabilities over all valid bitstrings, but rather each bitstring's sampling probability is re-weighted individually by its cost. With this method, one can utilize generative models to learn desired bitstrings for optimization-based tasks that correspond to real-world, relevant problems. A model is assessed for its quality-based generalization capabilities by looking at its \emph{Utility} $U$,
computed as
\begin{equation}
    U = \langle c(\boldsymbol{x}) \rangle_{\boldsymbol{x}\in P_5},
    \label{eq:utility}
\end{equation}
where $c(\boldsymbol{x})$ is the corresponding cost of a sample $\boldsymbol{x}$. $P_5$ corresponds to the set of samples with the lowest $5\%$ costs of unseen and valid queries. One is therefore looking to determine if the utility of the model is lower when trained on a uniform distribution rather than on a reweighted distribution, and if the utility of the model is lower than that of each respective training distribution. The former question tells us if the model is able to learn the `reweighting bias' introduced in the training set, and the latter tells us if the model is able to go beyond the samples in the training set and optimize further for generalized high quality samples. We note that other quality-based metric variations can be introduced as dependent on the objective of the task. 

\section{Results}\label{s:results}

We showcase our results on the validity-based and quality-based generalization capabilities of QCBMs. After providing the details of our simulation, we utilize the framework described in \secref{sec2:eval_gen} to monitor the $(F, \tilde{R}, \tilde{C})$ values of the QCBM throughout training with various circuit depths. Taking the metric values from the last training iteration, we then compare the models across $\epsilon$ in order to investigate the minimum data requirements to obtain high quality generalization performance. 
Lastly, we investigate the QCBM's ability to go beyond the validity generalization, and to generate high-quality (low cost) samples post learning from a re-weighted training set. 

\subsection{Simulation Details}\label{sec3:sim_details}

For all numerical experiments in \secref{sec3:vbased}, we train a 12-qubit QCBM circuit with a line topology to learn a cardinality-constrained target distribution $P(x)$. This dataset has a solution space $\mathcal{S}$ defined by bitstrings that have a specific number $k$ of 1s (e.g. `10001011' for $k = 4$). Thus the target distribution is uniform over the solution space: 

\begin{equation}\label{data_prob}
P(x) = \frac{1} {|\mathcal{S}|} \quad \forall x \in \mathcal{S}.
\end{equation}

The definition of such target distribution is only formal: in reality, it is not needed to have \textit{a priori} knowledge of $|\mathcal{S}|$ for our metrics to be determined. The coverage metric (\eqref{coverage}) seems to require such knowledge. This requirement can be met when addressing benchmarking instances, such as the ones investigated in this work. However, in real-world cases, one is usually interested in comparing models (either different models or two different versions of the same model), hence one can simply compute their $C$ ratios, which mitigates the fact that  $|\mathcal{S}|$ is not known. Additionally, if one is aiming at estimating the coverage of a standalone model, one could simply use the asymptotic limit of the normalized coverage, given by ~\eqref{coverage_asymptotic}, which quantifies the total number of unique unseen and valid samples over the total number of queries and does not require knowledge of $|\mathcal{S}|$. This value would provide a sufficient estimation with regards to the number of unique values covered from the solution space that were not seen during training. The important notion in this framework is having a target distribution which is uniform and whose support is given by samples in a well-defined valid sector. Although in the case of the cardinality-constrained data set $|\mathcal{S}|$ can be estimated exactly, there are several real-world instances where it is easy to determine whether a sample belongs to the valid space, but it is intractable to determine \textit{a priori} the size of the support. Examples of this class of problems can be found in Appendix 6 of Garey and Johnson's comprehensive book on NP-Complete problems~\cite{garey79}. For example, the \emph{zero-one integer programming} problem described as MP1 in Appendix 6 is an NP-Complete problem where it is easy to check whether a given sample satisfies the constraints, but where it is intractable to find the whole set of bitstrings which satisfy the constraints.

We note that for this specific study, it is important to have a large $|\mathcal{S}|$ for assessing generalization, so that we can have appropriately sized training sets when spanning over $\epsilon$. As such, we choose $k = 6$ for all runs, where $|\mathcal{S}| = {\binom{12}{6}} = 924$. The circuits are trained via a gradient-free Covariance Matrix Adaptation Evolution Strategy (CMA-ES) optimizer \cite{hansen2016cma} with a NLL loss function on the Qulacs quantum simulator \cite{qulacs2021}. While we optimize with NLL, we display the cost values in the form of the KL Divergence, as it is easier to visualize the success of the training process. Indeed, we know that for $P_{\text{train}} = P_{\text{model}}$, we have $\text{KL} = 0$. Each circuit is initialized randomly, and the maximum number of training iterations is $i_{\text{Max}} = 10,000$. Lastly, $10,000$ samples are generated for the generalization evaluation procedure. 

\begin{figure*}
\includegraphics[width=\linewidth]{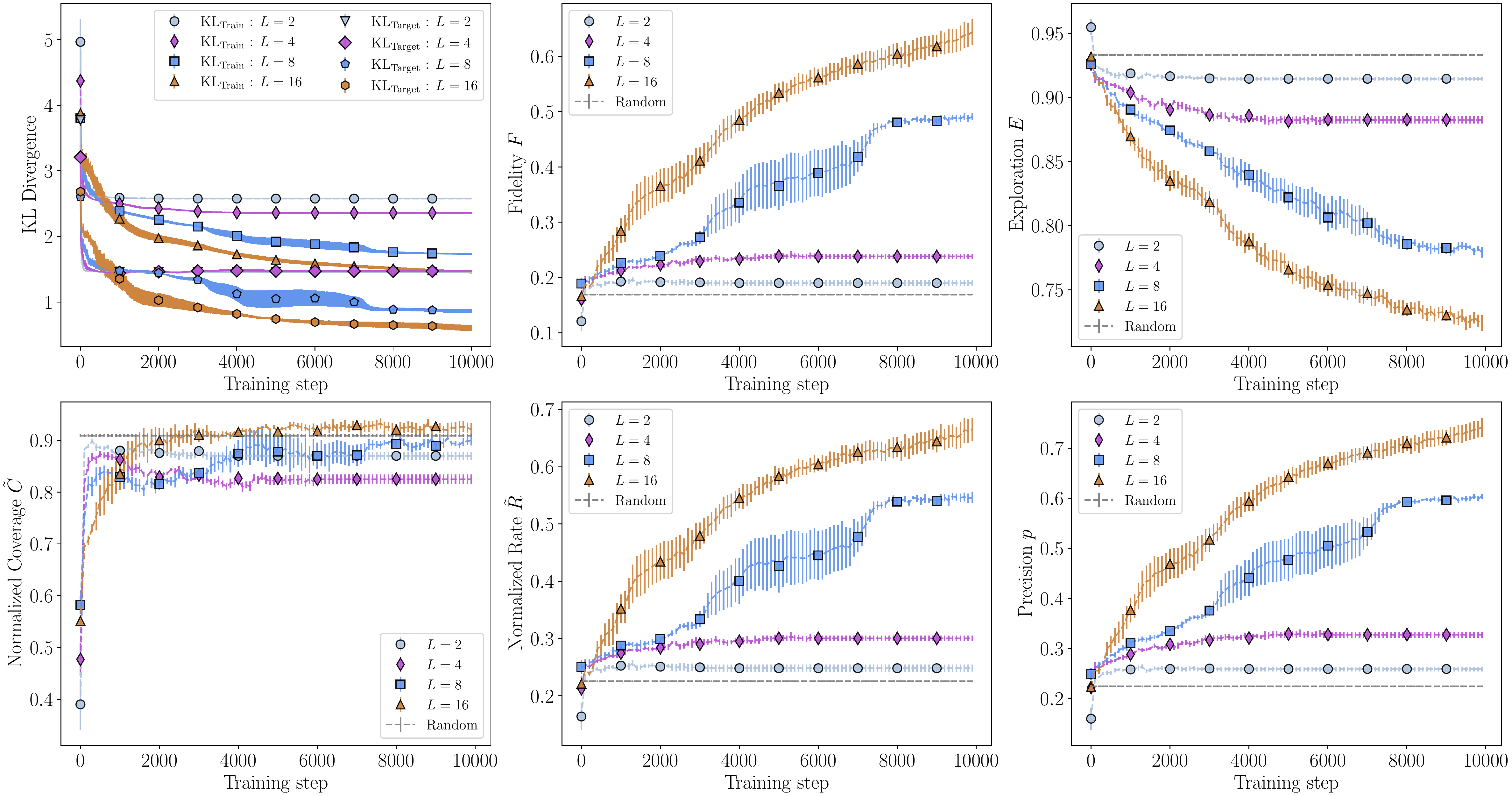}
\caption{\textbf{The QCBM's validity-based generalization performance throughout training across various circuit depths.} For each model with a different circuit depth $L \in \{2, 4, 8, 16\}$ and for $\epsilon = 0.3$, we show various metrics per training iteration including: the KL divergence of the model distribution relative to both the training and the target distribution (top left), the fidelity $F$ (top middle), the exploration $E$ (top right), the normalized coverage $\tilde{C}$ (bottom left), the normalized rate $\tilde{R}$ (bottom middle), and the precision $p$ (bottom right). Note that these are average values over 5 independent trainings, where the error bars are computed by $\sigma / \sqrt{5}$. We see that the generalization performance increases throughout training, and that while for the majority of metrics all models are able to beat the random search baseline, the generalization performance is best for the deepest circuit model.
}
\label{fig:spanning_layers}
\end{figure*}

To assess the QCBM's quality-based generalization in \secref{sec3:quality_gen}, we introduce a re-weighted \emph{Evens} dataset, where the solution space $\mathcal{S}$ is defined by bitstrings containing an even number of 1s (e.g. `01010011'), and a cost is assigned to each sample that quantifies the sample's degree of \emph{separation} $\gamma$. We define \emph{separation} as the largest bit-separation between 1s in the bitstring (e.g. $\gamma(\text{`11010001'}) = 4$). As we would like to optimize for the samples with the lowest cost, we focus on the negative separation $c = -{\gamma}$. We utilize this cost to reweight the training distribution via a softmax function on the training set bitstrings, namely: 
\begin{equation}\label{biased_prob}
P_{b}(x) = \frac{\exp\left(-\beta c(x)\right)}{\sum_{i = 1}^{|\mathcal{D}_{\text{Train}}|}\exp\left(-\beta c(x_i) \right) }.
\end{equation}

Following Ref.~\cite{alcazar2021enhancing}, we set $\beta = \beta_1 \equiv 1/T$ where $T$ is the standard deviation of the costs that is interpreted as a `temperature constant'~\cite{Alcazar2020ClvsQuant}. Note that by adjusting $\beta$, one can tune the degree of reweighting introduced into the uniform distribution: for instance, choosing $\beta = \beta_2 \equiv 2/T$ increases the impact of the reweighting procedure. The artifact of reweighting the target distribution allows one to determine whether or not the model is able to learn this `bias' induced by the bitstring costs in addition to the validity constraint.

The only additional change from the numerical experiments in \secref{sec3:vbased} is that we use an all-to-all $L =2$ ansatz rather than a line topology. In our numerical experiments, we found that using an all-to-all topology provided better validity-based generalization performances compared to the line topology. This observation is possibly related to a trainability issue of the latter topology at the number of layers needed to describe the reweighted dataset, since the uniform even distribution can be constructed exactly using a two-layered QCBM with a line topology~\cite{Exact_constructions_paper}. In general, for an arbitrary real-world dataset or more generic cases where one does not necessarily have an intuition of the type of circuit ansatz which might be suitable for the data, one needs to treat the exploration of the ansatz as an additional hyperparameter to be adjusted. Additionally, we use a fixed value of $\epsilon$, with $\epsilon = 0.1$. As the \emph{Evens} dataset is easier for the QCBM to learn, we use fewer layers and are thus able to avoid having too many parameters by implementing the all-to-all ansatz. We also remark that the choice of all-to-all compared to line topology on the \emph{Evens} dataset improves trainability significantly even though the line topology is sufficient to represent this state. 

Note that because the Evens solution space for the quality-based generalization assessment is much larger, 10\% of the solution space accounts for a similar number of samples as used in the validity-based investigation. This might have helped in seeing a comparable performance with less percentage of data than the former dataset, although in reality the properties of each distribution to be learned can play a significant role and this data efficiency capability needs to be studied on a case-by-case basis. For the validity-based generalization demonstration in this work, we emphasize that the model is able to learn from a few training data, and as such, we span over $\epsilon$ intentionally. In a practical context, however, there is no need to know this percentage since all the metrics, quality and validity-based, can be computed without its knowledge.

\subsection{Validity-Based Generalization}\label{sec3:vbased}

\subsubsection{Increasing Expressivity}\label{sec3:overfitting}

\begin{figure*}
\includegraphics[width=\linewidth]{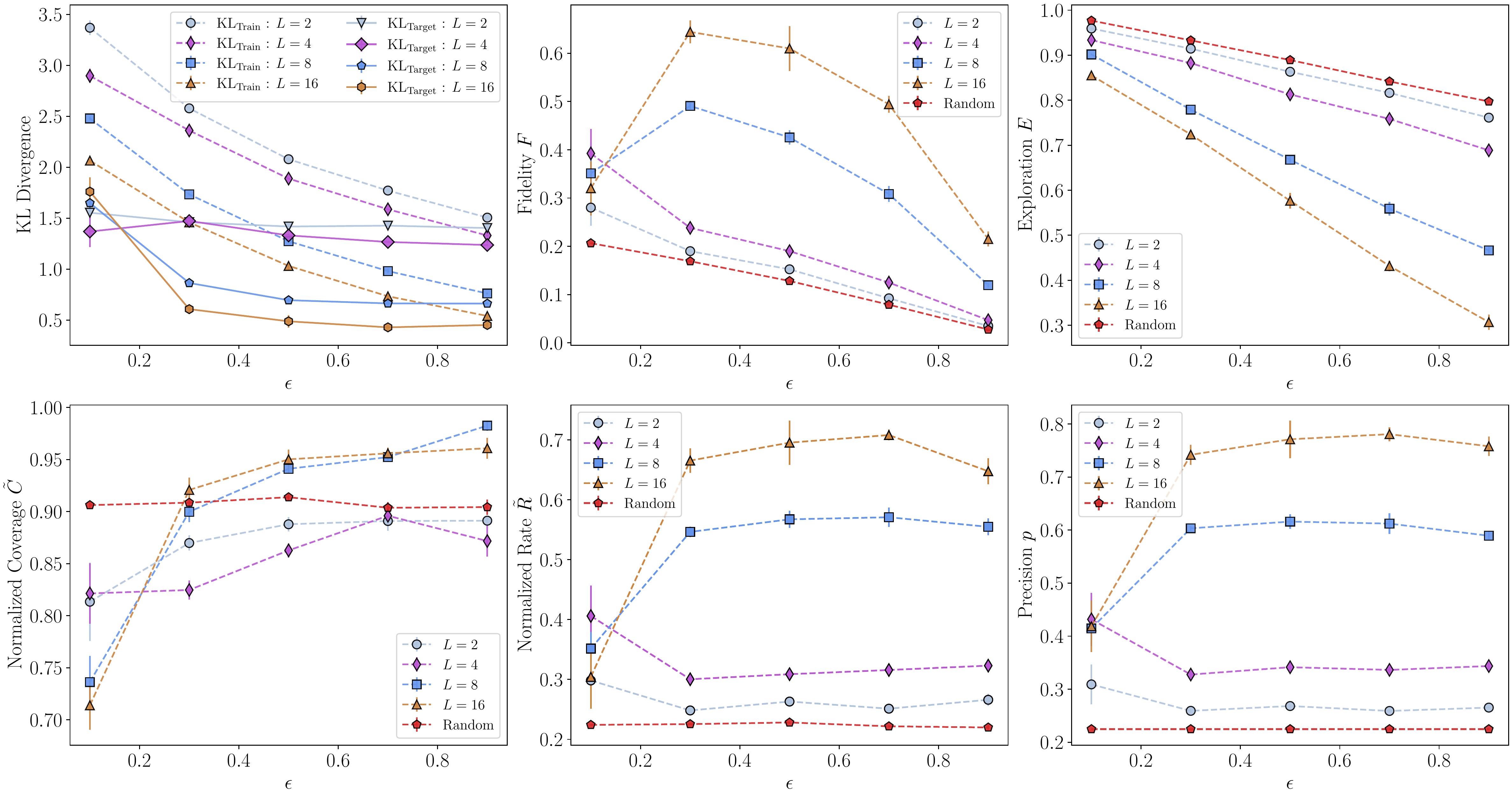}
\caption{\textbf{The QCBM's validity-based generalization performance on various training dataset sizes.} For each model with a different circuit depth $L \in \{2, 4, 8, 16\}$, we show various metrics across $\epsilon \in \{0.1, 0.3, 0.5, 0.7, 0.9\}$ values at the last iteration in training: the KL divergence of the model distribution relative to both the training and the target distribution
(top left), the fidelity $F$ (top middle), the exploration $E$ (top right), the normalized coverage $\tilde{C}$ (bottom left), the normalized rate $\tilde{R}$ (bottom middle), and the precision $p$ (bottom right). Note that these are average values over 5 independent trainings, where the error bars are computed as $\sigma / \sqrt{5}$. We see that the generalization performance increases for increasing circuit depth across all $\epsilon$ values. We see that for $L=16$, with as few as $30\%$ of the solution space used for training, the QCBM is able to exhibit great generalization performance on average: $(F, \tilde{R}, \tilde{C}) = (0.65, 0.67, 0.92)$, compared to that of the random baseline (in red): $(0.17, 0.23, 0.91)$.}
\label{fig:scaling_epsilon}
\end{figure*}

We show an example of the validity-based generalization for QCBM models trained with various circuit depths $L \in \{2, 4, 8, 16\}$. We only ran experiments with an even number of layers, where, as explained before, every layer of single qubit gates is followed by a layer of entangling gates. For the expressivity study presented here, we utilize only $30\%$ ($\epsilon = 0.3$) of the solution space in all runs, where the training data is uniformly sampled from all valid data (i.e., the solution space). In \figref{fig:scaling_epsilon}, we demonstrate that this is an adequate value for seeing good generalization performance. We run 5 independent trainings of each model, and plot the result averages with error bars in \figref{fig:spanning_layers}. 

Across all results, we see that the deepest QCBM with $L = 16$ produces the best performance. Indeed, we see a direct correlation between increasing the number of layers, or the expressivity of the circuit, and a higher generalization performance quantified by the validity-based metrics. For shallower circuits with $L \in \{2, 4\}$, we barely see any increase in generalization performance across the average $(F, \tilde{R}, \tilde{C})$ values throughout training. Additionally, we see very little training in general as evident from looking at both the corresponding KL divergence, exploration, and precision trends. These models barely beat the random baseline~\footnote{The random baseline for each metric is computed from samples randomly drawn from the set of $2^N$ possible bitstrings~\cite{gili2022evaluating}.}. After increasing the expressivity to $ L \in \{8, 16\}$, we begin to see the model learn. The average $(F, \tilde{R}, \tilde{C})$ values steadily increase, where we see that the $L=16$ model is able to reach the average values $(0.65, 0.67, 0.92)$, thus achieving high quality performance well above the average random baseline $(0.17, 0.23, 0.91)$. These metrics clearly show that the $L=16$ QCBM is producing more unseen and valid samples with each new iteration step, and is learning the valid samples, distinguishing them from the noise. At the same time, the model's total exploration decreases as expected, as it begins to produce both valid samples from inside and outside of the training set.  \tableref{table:valid_average} displays individual values for the different metrics and \tableref{table:random_baseline} includes values for the random baseline. 

\begin{figure*}
\includegraphics[width=\linewidth]{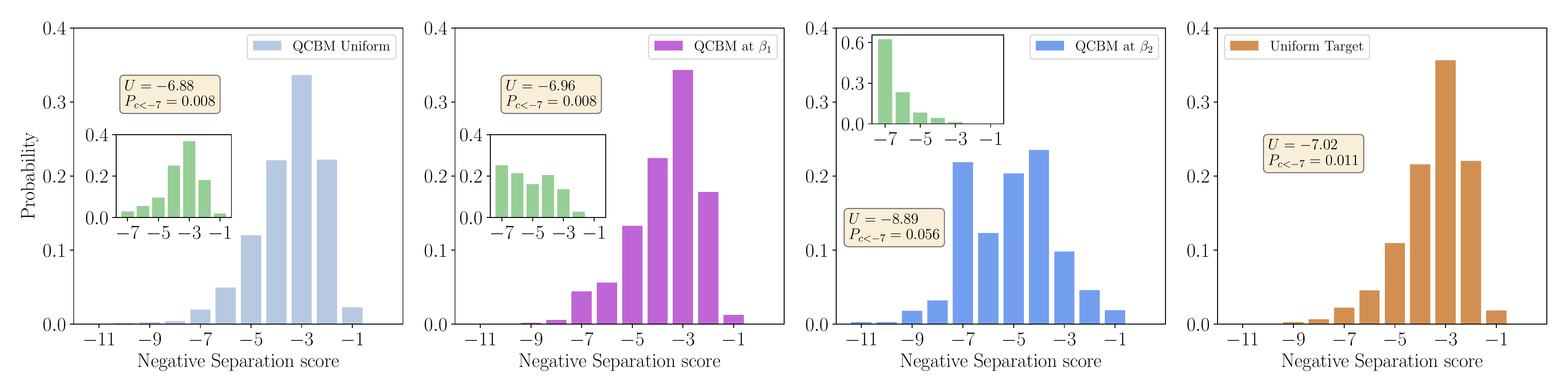}
\caption{\textbf{QCBM median output probability distributions when learning uniform vs. re-weighted datasets over $15$ independent runs.} Each model is trained on a set of bitstrings in the \emph{Evens} dataset, where each bitstring is mapped to its respective negative separation score, and to each training set a different degree of reweighting is applied. Note that we showcase the typical run across all models, as computed by the median $(F, \tilde{R}, \tilde{C})$ scores. We also note that the number of queries used to plot the histograms is $10000$, except for the (green) histograms in the insets which correspond to the 204 samples used as the training sets ($\epsilon=0.1$) for each degree of reweighting. Starting from the left, we show: the QCBM output when trained on a uniform training set; the QCBM output when trained on a re-weighted training set via a softmax function at inverse temperature $\beta_1 = 1/T$; the QCBM output when trained on a re-weighted training set via a softmax function at inverse temperature $\beta_2 = 2/T$; and 10,000 samples from the underlying uniform target distribution. While it is possible to observe values up to $c < -7$ when we sample 10,000 from the target distribution, across all training sets containing only 204 samples, the models do not ``see" bitstrings with an associated $c < -7$. We highlight the quality-based generalization metric $U$ for each distribution as well as the cumulative probability for generating samples where $P_{c < -7}$. With these two values, we are able to detect that the QCBM is able to achieve a lower $U$ when we increase the level of reweighting in the training set, as well as a higher $P_{c < -7}$. Therefore, we see that the QCBM is able to effectively learn the `reweighting bias' and generalize to data with lower negative separation costs than that of the training set.}
\label{fig:quality}
\end{figure*}

While not always an attainable metric, due to the fact that the target distribution is typically unknown, one can compute the target KL Divergence $\text{KL}_{\text{Target}} = \text{KL}(P(x) || P_{\text{model}})$ and compare it to the usual KL Divergence, relative to the training set $\text{KL}_{\text{Train}} = \text{KL}(P_{\text{train}} || P_{\text{model}})$, in order to see if the QCBM's output distribution is closer to the target than the training distribution. We see in \figref{fig:spanning_layers} that for $L = 16$, we have $\text{KL}_{\text{Target}}$ is smaller compared to $\text{KL}_{\text{Train}}$, indicating that the model is not overfitting to the training set. Note that we take the definition of \emph{overfitting} to be \emph{memorization} of the training distribution. We highlight that the values for $\text{KL}_{\text{Train}}$ are quite high compared to $\text{KL}_{\text{Target}}$, and according to this metric alone we could infer that the model does not have good generalization. However, displaying this metric alongside the $(F, \tilde{R}, \tilde{C})$ values, we are able to see that good generalization performance occurs even when $\text{KL}_{\text{Train}}$ is higher than zero. These numerics further support that achieving a $\text{KL}_{\text{Train}} = 0$ does not necessarily mean the model is producing good generalization performance: this metric alone should not be utilized to assess a generative model unless one is interested only in its data-copying capability. 

We make a similar argument with the precision $p$ and the exploration $E$. While simply providing $p$ or $E$ would not be enough to quantify the model's generalization performance, observing these values alongside the $(F, \tilde{R}, \tilde{C})$ metrics give us a broader picture. As the fidelity increases, despite the exploration decreasing, we can infer that this is simply because the model is disregarding noisy unseen samples. Additionally, as the rate increases alongside increasing precision, we can infer that a decent-sized portion of the valid samples being generated are unseen. 

Interestingly, circuits with both two and four layers produce a similar $\text{KL}_{\text{Target}}$, and even though the 2-layers circuit displays a higher coverage, it contains more noise (less rate and fidelity than the 4-layers circuit). Although the models happen to have roughly the same $\text{KL}_{\text{Target}}$ value, our metrics allow one to disentangle the different generalization contributions and provide insights into the strengths and weaknesses of the models. This example also shows how judging performance just based on $\text{KL}_{\text{Train}}$ can be very misleading (with the 4-layers circuit largely outperforming the 2-layers one), while the generalization metrics are more faithful to the $\text{KL}_{\text{Target}}$.

\subsubsection{Reducing the Amount of Training Data }\label{sec3:training_data_span}

We investigate the effect of the size of the training set on the model's ability to learn the valid correlations and generalize accordingly. We span over various percentage portions of the solution space $\epsilon \in \{0.1, 0.3, 0.5, 0.7, 0.9\}$ to use in the training set. \figref{fig:scaling_epsilon} showcases the average generalization performance at the last training iteration across 5 independent trainings for circuit depths $L \in \{2, 4, 8, 16\}$. We see that the trend of increasing circuit depth for enhanced performance extends to multiple $\epsilon$ values, thus suggesting that having enough expressivity is crucial for generalization to occur in this data set. Independent of the number of layers, it seems that when $\epsilon = 0.1$, the amount of training data is too low for the model to properly learn. When $\epsilon = 0.9$, we enter the regime where there is too much training data, which leaves very little room for generalization. Overall, we see that the model achieves good performance with access to $30\%$ of the solution space during training. \tableref{table:valid_average} displays individual values for the different metrics and \tableref{table:random_baseline} includes values for the random baseline. 

For increasing $\epsilon$, we see that the average exploration $E$ decreases. Such behaviour is expected as we are artificially decreasing the number of unseen bitstrings that the model can generate by reducing the size of the unseen space. For the average $(F, \tilde{R}, \tilde{C})$ values, we note some interesting individual trends. $F$ decreases after $\epsilon = 0.3$, whereas $\tilde{R}$ and $\tilde{C}$, and even the precision $p$, continue to increase slightly before decreasing at $\epsilon = 0.9$. We believe this discrepancy in the fidelity is related to the increase in the number of noisy unseen samples relative to the number of valid samples that the model could generate as we increase $\epsilon$. This is similar to the effect we described in \secref{sec2:unsup_gen}.

The models achieve better performance than the random search baseline, with the only exception of coverage when the circuit depths are too low ($L = {2, 4}$). We see that $\text{KL}_{\text{Train}}$ tends to decrease with decreasing $\epsilon$, and never encounters a turning point or a plateau. However, although $\text{KL}_{\text{Target}}$ presents the same decreasing trend until $\epsilon =0.5$, it begins to plateau or change less dramatically until the largest value explored of $\epsilon =0.9$. These two trends confirm that the QCBM model is indeed generalizing and not memorizing the data set, as memorization would imply an increased value for $\text{KL}_{\text{Target}}$. The fact that $\text{KL}_{\text{Target}} < \text{KL}_{\text{Train}}$ is the expected behaviour for a model which is generalizing since this means that the learned model distribution is closer to the target distribution than to the distribution from the training set. The decreasing in the value of $\text{KL}_{\text{Train}}$ can be easily explained from this behaviour since the training set is closer as well to the target distribution for larger values of $\epsilon$. This in turns explain the closing of the gap between the values of $\text{KL}_{\text{Train}}$ and $\text{KL}_{\text{Target}}$.

In summary, we demonstrate that QCBMs are able to go beyond memorizing a training distribution, and learn valid features in an underlying target distribution. They exhibit strong validity-based generalization across $(F, \tilde{R}, \tilde{C})$ values. However, we note that the QCBM requires deeper circuits in order to obtain good generalization performance, and this may pose a challenge for obtaining good results on near-term hardware. We believe that understanding how the hardware's noise and connectivity may impact the generalization capabilities of these models is important future work.

\subsection{Quality-Based Generalization}\label{sec3:quality_gen}

We assess the QCBM's quality-based generalization capability to see if the model is able to go beyond learning validity features in a dataset and learn an adequately reweighted version of it. 
As defined in \secref{sec3:sim_details}, each bitstring in the \emph{Evens} dataset's solution space is given an associated score for its negative separation $ c = - \gamma$, such that we can create a re-weighted training distribution according to the softmax function defined in \eqref{biased_prob}. For 12 qubits, the negative separation cost is $ c \in [-11, -1]$ with $c(`100000000001') = -11$ and $c(`111111111111') = -1$ as the limiting cases\footnote{When there is only one or no `$1$' in a bitstring, the cost is assigned to be zero. Those are only marginal cases, and they do not affect our analysis.}.

In \figref{fig:quality}, we show the results of the typical behaviour, which we take to be the median value of the $(F, \tilde{R}, \tilde{C})$ results across 15 independent runs. Starting from the left, the first three distributions are the histograms of the valid samples generated after training the QCBM on the inset distributions (green). Across all $\epsilon = 0.1$ training sets (204 samples), the models do not see bitstrings with an associated $c < -7$. The distribution in the rightmost panel corresponds to 10,000 queries from the underlying uniform target distribution across all bitstrings in the \emph{Evens} dataset mapped to their negative separation score. We share the average metric values across $15$ independent runs in \tableref{table:quality_average} and for the typical run in \tableref{table:quality_median}. 

For all output distributions the precision and the validity-based generalization performance are very high, as supported by the results in \tableref{table:quality_median}. However, the distributions differ in their quality-based generalization capabilities, which we assess by computing the utility $U$ metric described in \eqref{eq:utility} and the cumulative probability for $c < -7$, denoted as $P_{c < -7}$. We introduce the latter metric as a method to ensure that we are not simply reaching the lowest scores due to over-sampling, as upon sampling enough, it is reasonable to assume that the model will reach bitstrings with scores $c < -7$. If the model's $P_{c < -7}$ is higher than that of the uniform target, we see this as a fair way to assess that the model has learned the `reweighting bias' introduced in the training set.

When trained on a uniform training set, the model generalizes to the underlying uniform target distribution and generates some missing low cost bitstrings such that it has a $U = -6.88$ and a $P_{c < -7} = 0.008$ for the median run.~\footnote{We take the median run to be the median value out of all $F + \tilde{R} + \tilde{C}$ values, where we use the sum as a combined optimal score.} When trained with a re-weighted softmax training set, we see that the utility decreases to $U = -6.96$, while $P_{c < -7}$ stays the same. However, when we turn up the degree of reweighting in the softmax function by halving the temperature constant $T$, we see that the model begins to learn beyond the validity-based features and on average generates more valid and unseen bistrings that have a lower associated cost than the minimum shown to the model through the training set. Remarkably, we see that the utility drops to $U = -8.89$ and the cumulative probability grows to $P_{c < -7} = 0.056$. These results support that the QCBM is able to effectively learn the `reweighting bias' in the distribution, provided it is strong enough, and generalize to data outside of the training set with lower associated costs than that of the training set. This non-negligible quality-based generalization performance indicates that the QCBM may be useful for practical tasks in optimization, where one is interested in generating samples of minimal cost, by using them in the Generator-Enhanced Optimization (GEO) framework proposed in Ref.~\cite{alcazar2021enhancing}. In future work, it would be beneficial to see if we can model higher-dimensional distributions with QCBMs for performing practical tasks with constrained optimization problems. 

\section{Outlook}\label{s:outlook}

In the pursuit of obtaining a better fundamental understanding of quantum generative models such as QCBMs for real-world tasks, it is imperative to investigate not only their data-copying capability, but also their learning capabilities via the assessment of generalization performance. In this work, we conduct the first formal quantitative investigation of the QCBM's ability to learn feature patterns by demonstrating its generalization performance on uniform and re-weighted distributions. Our results show that the QCBM exhibits good validity-based generalization performance when learning a desired feature in a cardinality-constrained dataset. Overall, we see generalization performance tradeoffs when tuning various parameters. For instance, we show that one needs to increase the circuit depth, thus enhancing its expressivity, to obtain improved generalization performance. Additionally, we see that the model only requires access to $30\%$ of the solution space to generalize well. We put forth these trends as an initial investigation into the QCBM's generalization capabilities. As we scale up the circuit width, we believe these trends will serve as a good starting point for testing. 

We see that the QCBM's generalization performance is highly dependent on the number of valid data given to the model during training, and as such, it is of interest to dive deeper and conduct a more thorough future investigation into the QCBM's scarce data regime. In general, obtaining an improved understanding on the selection of challenging distributions that QCBMs can learn well is an important part of future work. As we scale to larger quantum circuit models and aim to use QCBMs in a more practical sense, it is paramount to obtain a better understanding of what practical tasks could be handled well by QCBMs. Our quality-based generalization results indicate that QCBMs are a good candidate for constrained optimization tasks, as the model can learn the correlations behind valid samples with associated low costs. In future work, we intend to scale up the size of the QCBM models and assess their capabilities on practical optimization tasks, such as portfolio optimization, where generalization capabilities have been shown to be a valuable asset~\cite{gili2022evaluating, alcazar2021enhancing}. We foresee more elaborate training techniques for assessing the generalization performance of larger and deeper circuit sizes would be needed.

As this study is the first formal assessment of QCBM's generalization performance, we hope our investigations will open the door for tackling future questions regarding the interplay of trainability~\cite{Mcclear2018Barren,cerezo2021costfunction}, expressibility~\cite{holmes2022expressivity}, and generalization~\cite{gili2022evaluating,mmd,Hinsche2022} in these models. In our study, we fail to see signs of over-parameterization, and in accordance with classical ML, we feel that this behavior might come to light if we continued to increase the number of layers. It remains an open research question to understand over-parameterization in quantum models, especially in the context of quantum generative models. For example, it would be interesting to see if phenomena such as double-descent which is present in deep learning models~\cite{nakkiran2021deep} due to over-parametrization happens in QCBMs as well. Additionally, we believe it will be important to investigate the generalization performance of QCBMs on quantum hardware, and investigate how noise impacts the training resources required. In the pursuit of quantum advantage, it would be interesting to utilize the generalization framework in Ref.~\cite{gili2022evaluating} to compare the QCBM's learning and generalization capabilities with those of other classical state-of-the-art generative models. 

Most importantly, we aim to motivate the community to place importance on assessing the \emph{generalization} over \emph{memorization} capabilities when introducing a new quantum or classical generative model. We hope to see a shift in emphasis in the evaluation scheme towards prioritizing (or at least including) generalization performances.

\begin{acknowledgments} 
We would like to thank Manuel Rudolph and John Realpe Gomez for helpful discussions and insights. We also acknowledge Dax Enshan Koh for reviewing our manuscript. K.G. would like to recognize the Army Research Office (ARO) (contract W911NF-20-1-0038) and UKRI (MR/S03238X/1) for providing funding through a QuaCGR PhD Fellowship. M.H. would like to acknowledge funding from MITACS through MITACS Accelerate. Lastly, K.G. would like to acknowledge the places that inspired this work: Oxford, UK; Santorini, Greece; Muscat, Oman; Lisbon, Portugal; Berlin, Germany and Boston, Greensboro, USA. Thank you to all of the local people who became a part of the journey.
M.M. and K.G. would like to acknowledge places where collaboration for this work happened in person: Lake Como, Italy; Toronto, Canada; and Chicago, USA.

\end{acknowledgments}

\bibliography{quantum-ai.bib}

\onecolumngrid

\clearpage
\section{Appendix}\label{s:appendix}

We present supplemental data for the results displayed in \secref{s:results} for both validity and quality-based generalization performance.

\subsection{Validity-Based Generalization }\label{s:valid_app}

To complement our results in \secref{sec3:vbased}, we show the metrics' results for the last training iteration in \figref{fig:spanning_layers} and for the various $\epsilon$ values in \figref{fig:scaling_epsilon} in \tableref{table:valid_average}. All metric values are averages over 5 independent trainings, where the error is given by the standard deviation over the square-root of the number of runs. Similarly, the average random baseline values with corresponding errors are shown in \tableref{table:random_baseline}.

\begin{table}[htp]
\centering
\fontsize{8pt}{15pt}
\renewcommand{\arraystretch}{1.3}
\begin{tabular}{||>{\centering}p{3cm} | >{\centering}p{1.5cm} | >{\centering}p{1cm} |
>{\centering}p{1cm} |
>{\centering}p{1cm} |
>{\centering}p{1cm} |
>{\centering}p{1cm} |
>{\centering}p{1cm} ||}
\hline
\multicolumn{1}{||c|}{$L$} & \multicolumn{1}{|c|}{
$\epsilon$} & \multicolumn{1}{|c|}{
$F$} & \multicolumn{1}{|c|}{$\tilde{R}$} & \multicolumn{1}{|c|}{$\tilde{C}$} & \multicolumn{1}{|c|}{$p$} & \multicolumn{1}{|c|}{$\text{KL}_{\text{Target}}$} & \multicolumn{1}{|c||}{$\text{KL}_{\text{Train}}$}\\
\hline
\multicolumn{1}{||c|}{2} & \multicolumn{1}{|l|}{$0.1$} & \multicolumn{1}{|c|}{$0.28 \pm 0.04$} & \multicolumn{1}{|c|}{$0.29 \pm 0.04$} & \multicolumn{1}{|c|}{$0.81 \pm 0.04$} & \multicolumn{1}{|c|}{$0.31 \pm 0.04$} & \multicolumn{1}{|c|}{$1.6 \pm 0.2$} & \multicolumn{1}{|c||}{$3.37 \pm 0.07$} \\
\hline 
\multicolumn{1}{||c|}{2} & \multicolumn{1}{|c|}{$0.3$} & \multicolumn{1}{|c|}{$0.189 \pm 0.005$} & \multicolumn{1}{|c|}{$0.248 \pm 0.007$} & \multicolumn{1}{|c|}{$0.869 \pm 0.007$} & \multicolumn{1}{|c|}{$0.259 \pm 0.005$} & \multicolumn{1}{|c|}{$1.46 \pm 0.02$} & \multicolumn{1}{|c||}{$2.578 \pm 0.006$} \\
\hline
\multicolumn{1}{||c|}{2} & \multicolumn{1}{|c|}{$0.5$} & \multicolumn{1}{|c|}{$0.152 \pm 0.003$} & \multicolumn{1}{|c|}{$0.263 \pm 0.005$} & \multicolumn{1}{|c|}{$0.888 \pm 0.007$} & \multicolumn{1}{|c|}{$0.268 \pm 0.004$} & \multicolumn{1}{|c|}{$1.418 \pm 0.008$} & \multicolumn{1}{|c||}{$2.079 \pm 0.008$} \\
\hline
\multicolumn{1}{||c|}{2} & \multicolumn{1}{|c|}{$0.7$} & \multicolumn{1}{|c|}{$0.092 \pm 0.002$} & \multicolumn{1}{|c|}{$0.251 \pm 0.004$} & \multicolumn{1}{|c|}{$0.891 \pm 0.009$} & \multicolumn{1}{|c|}{$0.259 \pm 0.004$} & \multicolumn{1}{|c|}{$1.427 \pm 0.006$} & \multicolumn{1}{|c||}{$1.771 \pm 0.006$} \\
\hline
\multicolumn{1}{||c|}{2} & \multicolumn{1}{|c|}{$0.9$} & \multicolumn{1}{|c|}{$0.035 \pm 0.001$} & \multicolumn{1}{|c|}{$0.27 \pm 0.01$} & \multicolumn{1}{|c|}{$0.891 \pm 0.03$} & \multicolumn{1}{|c|}{$0.265 \pm 0.002$} & \multicolumn{1}{|c|}{$1.403 \pm 0.006$} & \multicolumn{1}{|c||}{$1.506 \pm 0.005$} \\
\hline
\multicolumn{1}{||c|}{4} & \multicolumn{1}{|c|}{$0.1$} & \multicolumn{1}{|c|}{$0.39 \pm 0.05$} & \multicolumn{1}{|c|}{$0.41 \pm 0.05$} & \multicolumn{1}{|c|}{$0.82 \pm 0.03$} & \multicolumn{1}{|c|}{$0.43 \pm 0.05$} & \multicolumn{1}{|c|}{$1.4 \pm 0.2$} & \multicolumn{1}{|c||}{$2.89 \pm 0.07$} \\
\hline
\multicolumn{1}{||c|}{4} & \multicolumn{1}{|c|}{$0.3$} & \multicolumn{1}{|c|}{$0.238 \pm 0.005$} & \multicolumn{1}{|c|}{$0.300 \pm 0.005$} & \multicolumn{1}{|c|}{$0.825 \pm 0.009$} & \multicolumn{1}{|c|}{$0.328 \pm 0.006$} & \multicolumn{1}{|c|}{$1.47 \pm 0.02$} & \multicolumn{1}{|c||}{$2.36 \pm 0.01$} \\

\hline
\multicolumn{1}{||c|}{4} & \multicolumn{1}{|c|}{$0.5$} & \multicolumn{1}{|c|}{$0.190 \pm 0.004$} & \multicolumn{1}{|c|}{$0.309 \pm 0.004$} & \multicolumn{1}{|c|}{$0.863 \pm 0.007$} & \multicolumn{1}{|c|}{$0.341 \pm 0.008$} & \multicolumn{1}{|c|}{$1.33 \pm 0.02$} & \multicolumn{1}{|c||}{$1.587 \pm 0.008$} \\
\hline
\multicolumn{1}{||c|}{4} & \multicolumn{1}{|c|}{$0.7$} & \multicolumn{1}{|c|}{$0.125 \pm 0.001$} & \multicolumn{1}{|c|}{$0.316 \pm 0.003$} & \multicolumn{1}{|c|}{$0.896 \pm 0.009$} & \multicolumn{1}{|c|}{$0.336 \pm 0.005$} & \multicolumn{1}{|c|}{$1.266 \pm 0.005$} & \multicolumn{1}{|c||}{$1.587 \pm 0.008$} \\
\hline
\multicolumn{1}{||c|}{4} & \multicolumn{1}{|c|}{$0.9$} & \multicolumn{1}{|c|}{$0.047 \pm 0.002$} & \multicolumn{1}{|c|}{$0.32\pm 0.01$} & \multicolumn{1}{|c|}{$0.87 \pm 0.01$} & \multicolumn{1}{|c|}{$0.344 \pm 0.006$} & \multicolumn{1}{|c|}{$1.237 \pm 0.009$} & \multicolumn{1}{|c||}{$1.329 \pm 0.009$} \\
\hline
\multicolumn{1}{||c|}{8} & \multicolumn{1}{|c|}{$0.1$} &\multicolumn{1}{|c|}{$0.35 \pm 0.03$} & \multicolumn{1}{|c|}{$0.35 \pm 0.03$} & \multicolumn{1}{|c|}{$0.74 \pm 0.03$} & \multicolumn{1}{|c|}{$0.41 \pm 0.03$} & \multicolumn{1}{|c|}{$1.6 \pm 0.1$} & \multicolumn{1}{|c||}{$2.48 \pm 0.02$} \\
\hline
\multicolumn{1}{||c|}{8} & \multicolumn{1}{|c|}{$0.3$} & \multicolumn{1}{|c|}{$0.491 \pm 0.007$} & \multicolumn{1}{|c|}{$0.546 \pm 0.009$} & \multicolumn{1}{|c|}{$0.89 \pm 0.01$} & \multicolumn{1}{|c|}{$0.603 \pm 0.005$} & \multicolumn{1}{|c|}{$0.86 \pm 0.03$} & \multicolumn{1}{|c||}{$1.73 \pm 0.01$} \\
\hline
\multicolumn{1}{||c|}{8} & \multicolumn{1}{|c|}{$0.5$} & \multicolumn{1}{|c|}{$0.43 \pm 0.01$} & \multicolumn{1}{|c|}{$0.57 \pm 0.01$} & \multicolumn{1}{|c|}{$0.941 \pm 0.004$} & \multicolumn{1}{|c|}{$0.62 \pm 0.01$} & \multicolumn{1}{|c|}{$0.69 \pm 0.02$} & \multicolumn{1}{|c||}{$1.27 \pm 0.02$} \\
\hline
\multicolumn{1}{||c|}{8} & \multicolumn{1}{|c|}{$0.7$} & \multicolumn{1}{|c|}{$0.31 \pm 0.02$} & \multicolumn{1}{|c|}{$0.57 \pm 0.02$} & \multicolumn{1}{|c|}{$0.952 \pm 0.004$} & \multicolumn{1}{|c|}{$0.61 \pm 0.02$} & \multicolumn{1}{|c|}{$0.66 \pm 0.02$} & \multicolumn{1}{|c||}{$0.98 \pm 0.02$} \\
\hline
\multicolumn{1}{||c|}{8} & \multicolumn{1}{|c|}{$0.9$} & \multicolumn{1}{|c|}{$0.119 \pm 0.005$} & \multicolumn{1}{|c|}{$0.55 \pm 0.01$} & \multicolumn{1}{|c|}{$0.983 \pm 0.005$} & \multicolumn{1}{|c|}{$0.589 \pm 0.009$} & \multicolumn{1}{|c|}{$0.662 \pm 0.009$} & \multicolumn{1}{|c||}{$0.761 \pm 0.009$} \\
\hline
\multicolumn{1}{||c|}{16} & \multicolumn{1}{|c|}{$0.1$} & \multicolumn{1}{|c|}{$0.32 \pm 0.06$} & \multicolumn{1}{|c|}{$0.30 \pm 0.05$} & \multicolumn{1}{|c|}{$0.71 \pm 0.02$} & \multicolumn{1}{|c|}{$0.42 \pm 0.05$} & \multicolumn{1}{|c|}{$1.2 \pm 0.1$} & \multicolumn{1}{|c||}{$2.07 \pm 0.02$} \\
\hline
\multicolumn{1}{||c|}{16} & \multicolumn{1}{|c|}{$0.3$} & \multicolumn{1}{|c|}{$0.64 \pm 0.02$} & \multicolumn{1}{|c|}{$0.67 \pm 0.02$} & \multicolumn{1}{|c|}{$0.92 \pm 0.01$} & \multicolumn{1}{|c|}{$0.74 \pm 0.02$} & \multicolumn{1}{|c|}{$0.61 \pm 0.05$} & \multicolumn{1}{|c||}{$1.46 \pm 0.02$} \\
\hline
\multicolumn{1}{||c|}{16} & \multicolumn{1}{|c|}{$0.5$} & \multicolumn{1}{|c|}{$0.61 \pm 0.05$} & \multicolumn{1}{|c|}{$0.69 \pm 0.04$} & \multicolumn{1}{|c|}{$0.950 \pm 0.009$} & \multicolumn{1}{|c|}{$0.78 \pm 0.04$} & \multicolumn{1}{|c|}{$0.46 \pm 0.06$} & \multicolumn{1}{|c||}{$1.03 \pm 0.04$} \\
\hline
\multicolumn{1}{||c|}{16} & \multicolumn{1}{|c|}{$0.7$} & \multicolumn{1}{|c|}{$0.49 \pm 0.02$} & \multicolumn{1}{|c|}{$0.708 \pm 0.009$} & \multicolumn{1}{|c|}{$0.956 \pm 0.005$} & \multicolumn{1}{|c|}{$0.78 \pm 0.01$} & \multicolumn{1}{|c|}{$0.43 \pm 0.02$} & \multicolumn{1}{|c||}{$0.73 \pm 0.02$} \\
\hline
\multicolumn{1}{||c|}{16} & \multicolumn{1}{|c|}{$0.9$} & \multicolumn{1}{|c|}{$0.22 \pm 0.02$} & \multicolumn{1}{|c|}{$0.65 \pm 0.02$} & \multicolumn{1}{|c|}{$0.96 \pm 0.01$} & \multicolumn{1}{|c|}{$0.76 \pm 0.02$} & \multicolumn{1}{|c|}{$0.45 \pm 0.03$} & \multicolumn{1}{|c||}{$0.54 \pm 0.03$} \\
\hline
\end{tabular} 
\caption{\textbf{Validity-based generalization values over various circuit depths and training set sizes.} The table lists the average $(F, \tilde{R}, \tilde{C})$ values, along with the precision $p$ and the KL divergences relative to the target and the training set, across 5 independent trainings with the associated error for each model. Note that these are all values computed after sampling from the fully trained model ($i_{\text{Max}} = 10\text{k}$) when learning the cardinality-constrained target distribution.}
\label{table:valid_average}
\end{table}

\begin{table}[htp]
\centering
\fontsize{8pt}{15pt}
\renewcommand{\arraystretch}{1.3}
\begin{tabular}{||>{\centering}p{3cm} | >{\centering}p{1.5cm} | 
>{\centering}p{1cm} |
>{\centering}p{1cm} |
>{\centering}p{1cm} |
>{\centering}p{1cm} ||}
\hline
\multicolumn{1}{||c|}{
$\epsilon$} & \multicolumn{1}{|c|}{
$F$} & \multicolumn{1}{c|}{$\tilde{R}$} & \multicolumn{1}{|c|}{$\tilde{C}$} & \multicolumn{1}{|c||}{$p$}  \\
\hline
\multicolumn{1}{||l|}{$0.1$} & \multicolumn{1}{|c|}{$0.206 \pm 0.002$} & \multicolumn{1}{|c|}{$0.224 \pm 0.002$} & \multicolumn{1}{|c|}{$0.906 \pm 0.003$} & \multicolumn{1}{|l||}{$0.225 \pm 0.001$} \\
\hline 
\multicolumn{1}{||l|}{$0.3$} & \multicolumn{1}{|c|}{$0.169 \pm 0.001$} & \multicolumn{1}{|c|}{$0.225 \pm 0.002$} & \multicolumn{1}{|c|}{$0.909 \pm 0.003$} & \multicolumn{1}{|l||}{$0.225 \pm 0.001$} \\
\hline 
\multicolumn{1}{||l|}{$0.5$} & \multicolumn{1}{|c|}{$0.128 \pm 0.001$} & \multicolumn{1}{|c|}{$0.228 \pm 0.002$} & \multicolumn{1}{|c|}{$0.914 \pm 0.003$} & \multicolumn{1}{|l||}{$0.225 \pm 0.001$} \\
\hline 
\multicolumn{1}{||l|}{$0.7$} & \multicolumn{1}{|c|}{$0.0790 \pm 0.0007$} & \multicolumn{1}{|c|}{$0.222 \pm 0.002$} & \multicolumn{1}{|c|}{$0.904 \pm 0.005$} & \multicolumn{1}{|l||}{$0.225 \pm 0.001$} \\
\hline 
\multicolumn{1}{||l|}{$0.9$} & \multicolumn{1}{|c|}{$0.0276 \pm 0.0005$} & \multicolumn{1}{|c|}{$0.219 \pm 0.004$} & \multicolumn{1}{|c|}{$0.904 \pm 0.007$} & \multicolumn{1}{|l||}{$0.225 \pm 0.001$} \\
\hline 
\end{tabular} 
\caption{\textbf{Random search baselines for the validity-based generalization metrics across various $\epsilon$ values.} We show the average $(F, \tilde{R}, \tilde{C})$ values alongside the precision $p$, when no training is conducted and samples are taken at random. The results are averaged from 5 independent runs, and displayed with their corresponding standard errors. }
\label{table:random_baseline}
\end{table}

\clearpage
\subsection{Quality-Based Generalization }\label{s:quality_app}

To complement our results in \secref{sec3:quality_gen}, we show the metrics' results for each output distribution in \figref{fig:quality}, containing different degrees of reweighting. Note that in \tableref{table:quality_average} all values are averages over 15 independent trainings, where their error is given by the standard deviation over the square-root of the number of runs. In \tableref{table:quality_median}, we show the median score of $F+\tilde{R}+\tilde{C}$ for each distribution, which directly corresponds to the model outputs displayed in \figref{fig:quality}. 

\begin{table}[htp]
\centering
\fontsize{8pt}{15pt}
\renewcommand{\arraystretch}{1.3}
\begin{tabular}{||>{\centering}p{3cm} | >{\centering}p{1.5cm} | >{\centering}p{1cm} |
>{\centering}p{1cm} |
>{\centering}p{1cm} |
>{\centering}p{1cm} ||}
\hline
\multicolumn{1}{||c|}{Distribution}  & \multicolumn{1}{|c|}{
$F$} & \multicolumn{1}{|c|}{$\tilde{R}$} & \multicolumn{1}{|c|}{$\tilde{C}$} & \multicolumn{1}{|c|}{$p$}  \\
\hline
\multicolumn{1}{||c|}{Uniform}  & \multicolumn{1}{|c|}{$0.74 \pm 0.07$} & \multicolumn{1}{|c|}{$0.69 \pm 0.06$} & \multicolumn{1}{|c|}{$0.76 \pm 0.02$} & \multicolumn{1}{|c|}{$0.77 \pm 0.06$}  \\
\hline 
\multicolumn{1}{||c|}{Reweighted $T$}  & \multicolumn{1}{|c|}{$0.74 \pm 0.07$} & \multicolumn{1}{|c|}{$0.69 \pm 0.06$} & \multicolumn{1}{|c|}{$0.74 \pm 0.02$} & \multicolumn{1}{|c|}{$0.77 \pm 0.06$} \\
\hline 
\multicolumn{1}{||c|}{Reweighted $T/2$}  & \multicolumn{1}{|c|}{$0.85 \pm 0.07$} & \multicolumn{1}{|c|}{$0.75 \pm 0.05$} & \multicolumn{1}{|c|}{$0.59 \pm 0.03$} & \multicolumn{1}{|c|}{$0.87 \pm 0.06$} \\
\hline 
\end{tabular} 
\caption{\textbf{Average validity-based generalization metrics when training on uniform vs reweighted \emph{Evens} distributions.} We show the average $(F, \tilde{R}, \tilde{C})$ values, along with the precision $p$ across 15 independent runs with the associated error for each model. Note that these are all values computed after sampling from the fully trained model ($i_{\text{Max}} = 10\text{k}$).  } 
\label{table:quality_average}
\end{table}

\begin{table}[htp]
\centering
\fontsize{8pt}{15pt}
\renewcommand{\arraystretch}{1.3}
\begin{tabular}{||>{\centering}p{3cm} | >{\centering}p{1.5cm} | >{\centering}p{1cm} |
>{\centering}p{1cm} |
>{\centering}p{1cm} |
>{\centering}p{1cm} |
>{\centering}p{1cm} |
>{\centering}p{1cm} ||}
\hline
\multicolumn{1}{||c|}{Distribution}  & \multicolumn{1}{|c|}{
$F$} & \multicolumn{1}{|c|}{$\tilde{R}$} & \multicolumn{1}{|c|}{$\tilde{C}$} & \multicolumn{1}{|c|}{$p$} & \multicolumn{1}{|c|}{$U$} & \multicolumn{1}{|c||}{$P_{c< -7}$} \\
\hline
\multicolumn{1}{||c|}{Uniform}  & \multicolumn{1}{|c|}{$0.99$} & \multicolumn{1}{|c|}{$0.89$} & \multicolumn{1}{|c|}{$0.81$} & \multicolumn{1}{|c|}{$0.99$} & \multicolumn{1}{|c|}{$-6.88$}& \multicolumn{1}{|l||}{$0.008$} \\
\hline 
\multicolumn{1}{||c|}{Reweighted $T$}  & \multicolumn{1}{|c|}{$1.0$} & \multicolumn{1}{|c|}{$0.91$} & \multicolumn{1}{|c|}{$0.79$} & \multicolumn{1}{|c|}{$1.0$} & \multicolumn{1}{|c|}{$-6.96$}& \multicolumn{1}{|l||}{$0.008$} \\
\hline 
\multicolumn{1}{||c|}{Reweighted $T/2$}  & \multicolumn{1}{|c|}{$1.0$} & \multicolumn{1}{|c|}{$0.82$} & \multicolumn{1}{|c|}{$0.79$} & \multicolumn{1}{|c|}{$1.0$} & \multicolumn{1}{|c|}{$-8.89$}& \multicolumn{1}{|l||}{$0.056$} \\
\hline 
\end{tabular} 
\caption{\textbf{Median validity-based generalization and quality-based metrics when training on uniform vs reweighted \emph{Evens} distributions.} We show the median $(F, \tilde{R}, \tilde{C})$ values, along with the precision $p$ for 15 independent runs. Additionally, we show $U$ and $P_{c<-7}$ values that correspond to the median $(F, \tilde{R}, \tilde{C})$ score. Note that these are all values computed after sampling from the fully trained model ($i_{\text{Max}} = 10\text{k}$).   } 
\label{table:quality_median}
\end{table}

\end{document}